\documentclass[prb, letterpaper, twocolumn, fleqn, floatfix, showpacs, showkeys ]{revtex4}

\newcommand{\ebg}{\epsilon_{\text{bg}}}
\newcommand{\esub}{\epsilon_{\text{sub}}}

\newcommand{\Fs}{F_{\text{s}}}
\newcommand{\Fi}{F_{\text{i}}}
\newcommand{\es}{\epsilon_{\text{s}}}
\newcommand{\etf}{\epsilon_{\text{TF}}}
\newcommand{\bv}{{\bf v}}

\newcommand{\br}{{\bf r}}
\newcommand{\bq}{{\bf q}}
\newcommand{\Phiind}{\Phi_\mathrm{ind}}
\newcommand{\FPhiind}{\FPhi_{\mathrm{ind}}}
\newcommand{\FPhi}{\widetilde{\Phi}}

\newcommand{\FPhiext}{\FPhi_{\mathrm{ext}}}
\newcommand{\Eind}{{\bf E}^>_\mathrm{ind}}

\usepackage{euscript, units, amsfonts, graphicx, dcolumn, fancyhdr}
\usepackage{times}
\usepackage{times, graphicx, units,mathrsfs,euscript,upgreek,amssymb,pifont}

\begin{document}

\title{Dynamic polarization of graphene by moving external charges: random phase approximation}

\author{K. F. Allison}
\affiliation{Department of Applied Mathematics, University of Waterloo, Waterloo, Ontario, Canada N2L 3G1 }
\author{D. Borka}
\author{I. Radovi\'{c}}
\author{Lj. Had\v{z}ievski}
\affiliation{VIN\v{C}A Institute of Nuclear Sciences, P.O.\ Box 522, 11001 Belgrade, Serbia  }
\author{Z. L. Mi\v{s}kovi\'{c}}
\email{zmiskovi@math.uwaterloo.ca} \affiliation{Department of Applied Mathematics, University of Waterloo, Waterloo, Ontario,
Canada N2L 3G1}

\date{November 2009}

\pacs{79.20.Rf, 34.50.Bw, 34.50.Dy}

\keywords{graphene, dielectric function, RPA, screening, stopping power, image interaction}

\begin{abstract}
We evaluate the stopping and image forces on a charged particle moving parallel to a doped sheet of graphene by using the
dielectric response formalism for graphene's $\pi$-electron bands in the random phase approximation (RPA). The forces are
presented as functions of the particle speed and the particle distance for a broad range of charge-carrier densities in
graphene. A detailed comparison with the results from a kinetic equation model reveal the importance of inter-band
single-particle excitations in the RPA model for high particle speeds. We also consider the effects of a finite gap between
graphene and a supporting substrate, as well as the effects of a finite damping rate that is included through the use of
Mermin's procedure. The damping rate is estimated from a tentative comparison of the Mermin loss function with a HREELS
experiment. In the limit of low particle speeds, several analytical results are obtained for the friction coefficient that show
an intricate relationship between the charge-carrier density, the damping rate, and the particle distance, which may be relevant
to surface processes and electrochemistry involving graphene.
\end{abstract}

\maketitle \thispagestyle{plain}

\begin{section}{Introduction}

The interactions of fast-moving charged particles with various
carbon nanostructures have been studied extensively in the context
of electron energy loss spectroscopy (EELS), typically using a
scanning transmission electron microscope (STEM) with incident
electron energies on the order of 100 keV. This technique has proven
to be a powerful tool for investigating the dynamic response of
carbon nanotubes \cite{Stephan_2002,Kramberger_2008} and, more
recently, graphene.\cite{Eberlein_2008} At such high incident
electron energies, these studies have revealed important properties
of both high-frequency $\sigma+\pi$ plasmon excitations ($\sim
15-30$ eV) and low-frequency $\pi$ plasmon excitations ($\sim 5$ eV)
in isolated single-wall carbon nanotubes \cite{Kramberger_2008} and
in free-standing, undoped graphene.\cite{Eberlein_2008} While the
studies on carbon nanotubes typically give plasmon dispersions at
large wavenumbers ($>0.1$ \AA$^{-1}$) in the axial direction,
\cite{Stephan_2002,Kramberger_2008} the study on graphene was
performed with an electron momentum transfer close to zero, although
it integrated over a significant in-plane component of the plasmon
wave vector.\cite{Eberlein_2008} In both cases, experimental data
was found to be in good agreement with \textit{ab initio}
calculations.\cite{Kramberger_2008,Eberlein_2008,Trevisanutto_2008}
In addition to \textit{ab initio} calculations, methods employing an
empirical dielectric tensor \cite{Taverna_2002} and a two-fluid,
two-dimensional (2D) hydrodynamic model for graphene
\cite{Mowbray_2006,Radovic_2007} have also been able to reproduce
the basic features of the $\sigma+\pi$ and $\pi$ plasmon excitations
in carbon nanostructures.

For lower-energy external moving charges, recent progress has been made in measuring the dispersion of low-frequency plasmon
excitations on solid surfaces using high-resolution reflection EELS (HREELS) with incident electron energies on the order of 10
eV. Such a measurement was performed on metallic surface-state electron bands, \cite{Diaconescu_2007} and the results were
interpreted theoretically using a dielectric-response model within the random phase approximation (RPA) that took into account
the typically parabolic band structures of the surface states. \cite{Diaconescu_2007,Alducin_2007,Silkin_2008} Furthermore, Liu
\textit{et al.} \cite{Liu_2008} have used HREELS to compare the low-frequency excitation spectra of doped graphene on a SiC
substrate with the spectra of a metallic monolayer on a semiconducting Si substrate. At such low incident electron energies, the
authors were able to measure the $\pi$ plasmon dispersion in a range of small wavenumbers ($<0.2$ \AA$^{-1}$) for a doped sheet
of graphene with a high charge-carrier density. \cite{Liu_2008} This HREELS experiment, which provides the wavenumber-resolved
spectra of low-frequency excitation modes in graphene with a high sensitivity to the doping level, \cite{Liu_2008} is more relevant
to the parameter space in the present work than the STEM-EELS experiments.

It is well appreciated that doping plays an immensely important role in graphene's conducting properties, for which electron
scattering on statically-screened charged impurities situated near graphene is one of the most important processes and is likely
responsible for the famed minimum conductivity in undoped graphene.
\cite{Novoselov_2005,Ando_2006,Peres_2007,Sarma_2007,PNAS_2007,Adam_2008,Chen_2008,Castro_2009} In this context, important
progress has been made in the development and use of the RPA dielectric function for low-energy excitations involving graphene's
$\pi$ electron bands in the approximation of linearized electron energy dispersion, which gives rise to the picture of massless
Dirac fermions (MDF). \cite{Shung_1986,Wunsch_2006,Hwang_2007,Barlas_2007} This progress has opened up a range of interesting
problems involving the interaction of graphene with external charges moving sufficiently slow that the MDF-RPA dielectric
response theory can be applied. While an obvious application of this theory would be to interpret the HREELS experiment
\cite{Liu_2008} on graphene, another interesting application would be to the study of slow, heavy particles moving near
graphene. This latter application of graphene's MDF-RPA dielectric response theory would be relevant to studies of chemisorption
of alkali-metal atoms, \cite{Khantha_2004} friction of migrating atoms and molecules \cite{Tomassone_1997, Dedkov_2005} moving
near graphene, and ion transport in aqueous solutions adjacent to graphene when top-gating with an electrolyte is implemented.
\cite{Bazant_2004} Moreover, one could explore the application of low-energy, ion-surface scattering techniques
\cite{Brongersma_2007} to graphene and other carbon nanostructures. There has also been recent interest in the directional
effects of ion interactions with graphene-based materials, such as low-energy ion channeling through carbon nanotubes
\cite{Miskovic_2007} and ion interactions with highly-oriented pyrolytic graphite, including implantation, \cite{Ramos_2001}
channeling, \cite{Yagi_2004} and ion-induced secondary electron emission from this target. \cite{Cernusca_2005} If applied to
graphene, most of the scattering configurations in these studies would involve impacts of slow, heavy particles under grazing
angles of incidence, and many interesting parallels may be found with Winter's experiments on the grazing scattering of ions and
atoms from solid surfaces. \cite{Winter_2002}

We therefore wish to study the application of the MDF-RPA model to charged particles moving parallel to a single layer of
supported graphene under gating conditions. In the wavenumber-frequency domain, $(q,\omega)$, the MDF-RPA model is applicable to
graphene's polarization modes if the conditions $q<2k_c$ and $\omega<2\varepsilon_c/\hbar$ are satisfied, where $k_c\approx
a^{-1}$ is a high-momentum cut-off (with lattice constant $a\approx$ 2.46 \AA) and $\varepsilon_c\approx$ 1 eV is a
high-frequency cut-off validating the approximation of linearized $\pi$ electron bands.
\cite{Wunsch_2006,Barlas_2007,Castro_2009} For a point charge moving parallel to graphene at a fixed distance $z_0$ and constant
speed $v$, the former condition will be satisfied only for distances $z_0>a$, and hence we may neglect both the size of the
particle and the size of the $\pi$ electron orbitals in graphene. The latter condition can be transformed into a restriction on
the particle speed by invoking the Bohr's adiabatic criterion and requiring that $v/z_0<2\varepsilon_c/\hbar$. It is clear that
with a gap on the order of 7 eV for graphene's $\sigma$ bands, particles moving at such slow speeds and large distances cannot
excite the high-energy modes involving graphene's $\sigma$ electrons.

Within the constraints of the MDF-RPA model, the main focus of this paper is on the stopping force and the dynamic image force
acting on an external charged particle. We note that the stopping force is equal to the negative of the stopping power, which is
defined as the energy loss of the external particle per unit length along its trajectory. \cite{Winter_2002} Meanwhile, the
image force is a conservative force \cite{Gumbs_2009} that can strongly deflect a particle's trajectory, especially for low
particle speeds and/or small angles of incidence upon the target's surface. \cite{Winter_2002} This was demonstrated not just
for electron interactions with solid surfaces, \cite{Echenique_1985,Lambin_1987,Inaoka_2001} but also for ion \cite{Song_2003}
and molecule \cite{Song_2005} grazing scattering from solid surfaces and ion \cite{Zhou_2005,Borka_2006} and molecule
\cite{Zhou_2006,Mowbray_2007} channeling through carbon nanotubes. For example, in Ref.\cite{Song_2003} it was shown that both
the stopping and image forces must be treated in a self-consistent manner in order to model ion trajectories and obtain ion
energy losses that agree well with experiment results for the grazing scattering of slow, highly-charged ions on various
surfaces. \cite{Fritz_1996,Juaristi_1999} A discussion of the stopping and image forces in the MDF-RPA model is therefore
relevant to the current literature.

In our previous work, \cite{Radovic_2008} we have calculated the stopping and image forces on charged particles moving above
graphene by assuming a high equilibrium density, $n$, of charge carriers in graphene and using a kinetic (Vlasov) equation to
describe the response of graphene's $\pi$ bands within the linearized electron energy dispersion approximation.
\cite{Radovic_2008,Ryzhii_2007} This semi-classical kinetic equation (SCKE) model gave a relatively simple dielectric function
for graphene that accurately described the thermal effect on plasmon dispersion \cite{Ryzhii_2007,Vafek_2006} and allowed us to
analyze the contributions of plasmon excitations and low-energy intra-band single-particle excitations (SPEs) to the stopping
and image forces. \cite{Radovic_2008} However, it remained unclear how large the density must be to validate the semi-classical
model and, more importantly, what effect the inter-band SPEs that lie beyond the capability of the SCKE model have. Therefore,
the first goal of this paper is to determine the conditions under which the SCKE model is applicable at zero temperature by
comparing the stopping and image forces obtained using the SCKE dielectric function \cite{Radovic_2008} with those obtained
using the MDF-RPA dielectric function. \cite{Wunsch_2006,Hwang_2007,Barlas_2007} Furthermore, since we have found in Ref.\
\cite{Radovic_2008} that a finite gap between graphene and the substrate strongly affects both forces in the SCKE model, the
second goal of this paper is to examine the effect of a finite gap in the MDF-RPA model. We note that the issue of a finite gap
has become more important as increasingly diverse dielectric environments for graphene are studied. \cite{Jang_2008}

Although we consider the MDF-RPA dielectric function to be a basic,
parameter-free model that provides an adequate description of the
inter-band SPEs in graphene, the model nevertheless has its
shortcomings. For example, it ignores the local-field effects (LFE)
due to electron-electron correlations
\cite{Trevisanutto_2008,Kramberger_2008} and assigns an infinitely
long lifetime to the electron excitations. The latter deficiency is
often rectified in \textit{ab initio} studies by applying a finite
broadening, on the order of 0.5 eV, to the frequency domain for
calculations of the loss function. \cite{Eberlein_2008} In a similar
way, one can introduce a finite relaxation time, or decay (damping)
rate, $\gamma$, to the MDF-RPA dielectric function for graphene
using Mermin's procedure. \cite{Mermin_1970,Qaiumzadeh_2008} Since
there are many scattering processes that can give rise to a finite
lifetime of the excited $\pi$ electrons in graphene, an accurate
determination of $\gamma$ still presents a
challenge.\cite{Hwang_2008,Qaiumzadeh_2008} Therefore, the third
goal of this paper is to treat $\gamma$ as an empirical parameter
and investigate the effects of a finite damping rate on the stopping
and image forces calculated with a MDF-RPA dielectric function
modified by the Mermin procedure. This dielectric function,
hereafter referred to as the Mermin dielectric function, requires a
careful extension of the MDF-RPA dielectric function derived for
$\gamma=0$ in Refs.\cite{Wunsch_2006,Hwang_2007} to finite $\gamma$.
Details of the Mermin dielectric function are given in Appendix A.

The parameters of primary interest in this study are therefore the equilibrium density of charge carriers in graphene, $n$, the
graphene-substrate gap height, $h$, and the damping rate, $\gamma$. The equilibrium density is particularly important because it
determines the Fermi momentum of graphene's $\pi$-electron band, $k_F=\sqrt{\pi n}$, and the corresponding Fermi energy,
$\varepsilon_F=\hbar v_Fk_F$, where $v_F\approx c/300$ is the Fermi speed of the linearized $\pi$ band and $c$ is the speed of
light in free space. In the case of intrinsic graphene ($n=0$), the Fermi energy coincides with the Dirac point,
$\varepsilon_F=0$. In this paper, we consider a wide range of densities $n\ge0$, expressed as a multiple of the base value
$n_0=10^{11}$ cm$^{-2}$, under the conditions $k_F<k_c$ and $\varepsilon_F<\varepsilon_c$.

The rest of the paper is outlined as follows. In section II, we present a theoretical derivation of the interaction of a general
charge distribution with a layer of supported graphene. This derivation motivates the definition of the stopping and image
forces for a point charge. In section III, we compare the stopping and image forces in the MDF-RPA and SCKE models for the
simple case of free graphene and a vanishing damping rate to determine the range of densities for which the SCKE model is valid.
We then focus on the MDF-RPA model with a vanishing damping rate, and in section IV we investigate the effects of a finite gap
between graphene and a SiO$_2$ substrate. In the simplified case of a zero gap, we provide analytic expressions for the stopping
and image forces for intrinsic graphene and low particle speeds. Finally, in section V we consider the MDF-RPA model with a
finite damping rate. After comparing the MDF-RPA model with experimental data to estimate the value of the damping rate, we
consider the effects of the damping rate on the stopping and image forces with a special focus on the stopping force at low
particle speeds. Note that we use Gaussian electrostatic units.

\end{section}

\begin{section}{Basic theory}

We first give a brief generalization of the formalism developed in Ref.\cite{Radovic_2008} to the case of a charge distribution
with density $\rho_\mathrm{ext}(\br,z,t)$. We assume that the charge distribution moves along a classical trajectory in a
Cartesian system with graphene placed in the $z=0$ plane and with coordinates $\br=\{x,y\}$ in the graphene plane. In keeping
with the reflection geometry of ion-surface grazing scattering, \cite{Winter_2002} we assume that the external charge
distribution remains localized in the region $z>0$ above graphene while a semi-infinite substrate occupies the region $z<-h$
below graphene. Following Ref.\cite{Radovic_2008}, we can express the induced potential $\Phiind(\br,z,t)$ in the region above
graphene using the Fourier transform ($\br\rightarrow\bq$ and $t\rightarrow\omega$) as
\begin{eqnarray}
\FPhiind^>({\bq},z,\omega)= \left[\frac{1}{\epsilon(\bq,\omega)}-1\right]\FPhiext^0({\bq},\omega)\,\mathrm{e}^{-qz} ,
\label{Phiind}
\end{eqnarray}
where $\epsilon(\bq,\omega)$ is the dielectric function of the combined graphene-substrate system and $\FPhiext^0(\bq,\omega)$
is the Fourier transform of the external potential evaluated at the graphene plane, $z=0$. The dielectric function of the system
can be written as
\begin{eqnarray}
\epsilon(\bq,\omega)=\ebg(q)+\frac{2\pi e^2}{q}\Pi(\bq,\omega), \label{eps}
\end{eqnarray}
where $\Pi(\bq,\omega)$ is the polarization function for free graphene and $\ebg(q)$ is the effective background dielectric
function, which is expressed in terms of the substrate dielectric constant $\esub$ as \cite{Radovic_2008}
\begin{equation}
\label{eq:ebg} \ebg(q)=\frac{\esub+1}{2}\,\frac{1+\coth(qh)}{\esub+\coth(qh)}.
\end{equation}

We note that, instead of dielectric constant $\esub$, one may use a frequency dependent substrate bulk dielectric function,
$\esub(\omega)$, in order to include the effects of coupling between graphene's $\pi$ electrons and either the surface phonon
modes in a strongly polar insulating substrate or the surface plasmon modes in a metallic substrate under the local
approximation. \cite{Fischetti_2001} In the former case, which includes a substrate with a single transverse optical (TO) phonon
mode at frequency $\omega_\mathrm{TO}$, one may use a dielectric function of the form \cite{Fratini_2008}
\begin{equation}
\label{eq:esub} \esub(\omega)=\epsilon_\infty+\left( \epsilon_s-\epsilon_\infty
\right)\frac{\omega_\mathrm{TO}^2}{\omega_\mathrm{TO}^2-\omega\!\left(\omega+i\gamma_\mathrm{TO}\right)},
\end{equation}
where $\epsilon_s=\esub(0)$ and $\epsilon_\infty=\lim_{\omega\rightarrow\infty}\esub(\omega)$ are the static and high-frequency
dielectric constants of the substrate, respectively. In the latter case, which includes the high-frequency response of a metal,
one may use the Drude dielectric function $\esub(\omega)=1-\omega_p^2/\!\left[\,\omega\!\left(\omega+i\gamma_{p}\right)\right]$
with a plasma frequency $\omega_p$ and a damping rate $\gamma_{p}$.

We limit the focus of this work to an insulating substrate in the static mode with a dielectric constant $\esub=\epsilon_s$, but
we allow for an arbitrary gap $h$ between graphene and the substrate. We note, however, that it is common in the literature to
assume a zero gap, \cite{Wunsch_2006,Hwang_2007,PNAS_2007,Adam_2008} for which Eq.\ (\ref{eq:ebg}) gives an effective background
dielectric constant $\ebg^0=(\es+1)/2$. In this case, a simple description of the screening of electron-electron interactions in
graphene can be quantified by the Wigner-Seitz radius, $r_s=e^2/(\ebg^0\hbar v_F)$, \cite{Hwang_2007} and free graphene can be
recovered by setting $\es=1$, and hence $\ebg^0=1$. Results provided for $h=0$ are therefore slightly more general than results
provided for $h\rightarrow\infty$, which also characterizes free graphene by yielding $\ebg=1$ in Eq.\ (\ref{eq:ebg}).

Next, we write $\FPhiext^0(\bq,\omega)=\frac{2\pi}{q}S(\bq,\omega)$, where the external charge structure factor $S(\bq,\omega)$
is given by
\begin{eqnarray}
S(\bq,\omega)=\int\limits_{-\infty}^\infty dt\,\mathrm{e}^{i\omega t}\int d^2\br\,\mathrm{e}^{-i\bq\cdot\br}\int\limits_0^\infty
dz\,\mathrm{e}^{-qz}\,\rho_\mathrm{ext}(\br,z,t).
 \label{Structure}
\end{eqnarray}
For a point charge $Ze$ moving parallel to graphene with velocity $\bv$ and at a fixed distance $z_0>0$, we find that
$S(\bq,\omega)=2\pi Ze\,\delta(\omega-\bq\cdot\bv)\,\mathrm{e}^{-qz_0}$. In this case, the induced electric field
$\Eind(\br,z,t)=-\nabla \Phiind^>(\br,z,t)$ can be written as

\begin{eqnarray}
\Eind(\br,z,t)=\frac{Ze}{2\pi} \times \nonumber \\
\int
d^2\bq\,\left(\,{\bf \hat{z}}-i{\bf
\hat{q}}\right)\,\mathrm{e}^{i\bq\cdot(\br-\bv
t)}\,\mathrm{e}^{-q\left(z+z_0\right)}\,\left[\frac{1}{\epsilon(\bq,\bq\cdot\bv)}-1\right],
 \label{Eind}
\end{eqnarray}
where ${\bf \hat{z}}$ is a unit vector perpendicular to graphene and ${\bf \hat{q}}=\bq/q$. For stopping and image forces
defined by $F_s=Ze\,{\bf \hat{v}}\cdot\Eind(\br\!=\!\bv t,z\!=\!z_0,t)$ and $F_i=Ze\,{\bf \hat{z}}\cdot\Eind(\br\!=\!\bv
t,z\!=\!z_0,t)$, respectively, where ${\bf \hat{v}}=\bv/v$, one obtains \cite{Radovic_2008}
\begin{eqnarray}
\Fs=\frac{2}{\pi}\frac{Z^2e^2}{v} \times \nonumber \\
\int_0^\infty
dq\,\mathrm{e}^{-2qz_0}\int_0^{qv}d\omega\,\frac{\omega}{\sqrt{q^2v^2-\omega^2}}\,\,
\Im\!\left[\frac{1}{\epsilon(q,\omega)}\right], \label{stopping2}
\end{eqnarray}
\begin{eqnarray}
\Fi=\frac{2}{\pi}Z^2e^2 \times \nonumber \\
\int_0^\infty
dq\,q\,\mathrm{e}^{-2qz_0}\int_0^{qv}\frac{d\omega}{\sqrt{q^2v^2-\omega^2}}
\,\, \Re\!\left[\frac{1}{\epsilon(q,\omega)}-1\right].
\label{image2}
\end{eqnarray}
Note that we have used the symmetry properties of the MDF-RPA dielectric function $\epsilon(q,\omega)$ to simplify Eqs.\
(\ref{stopping2}) and (\ref{image2}).

For the comparison with the HREELS experiment \cite{Liu_2008} in section V, we also define the total energy of the external
charge reflected from graphene as
\begin{eqnarray}
E_\mathrm{loss}&=&-\int\limits_{-\infty}^\infty dt\int d^2\br\int\limits_{-\infty}^\infty dz\,\rho_\mathrm{ext}(\br,z,t)
\,\frac{\partial}{\partial t}\Phiind(\br,z,t) \nonumber \\
&=& \int\limits_0^\infty d\omega\,\omega\int
\frac{d^2\bq}{2\pi^2}\,\vert
S(\bq,\omega)\vert^2\,\Im\!\left[\frac{-1}{\epsilon(\bq,\omega)}\right]
.
 \label{Eloss}
\end{eqnarray}
For a point charge $Ze$ moving on a specular-reflection trajectory with $\rho_\mathrm{ext}(\br,z,t)=Ze\,\delta\!\left(
\br\!-\!\bv_{\parallel}t\right)\delta\!\left(z\!-\!v_{\perp}\vert t\vert\right)$, where $\bv_{\parallel}$ and $v_{\perp}$ are
the components of the particle velocity parallel and perpendicular to the graphene plane, respectively, the probability density
for exciting the mode with frequency $\omega$ and wavevector $\bq$ is \cite{Ibach_1982}
\begin{eqnarray}
P(\bq,\omega)=\frac{2}{\pi^2}\frac{(Ze)^2v_{\perp}^2q}{\left[\left(\omega-\bq\cdot\bv_{\parallel}\right)^2+
\left(qv_{\perp}\right)^2\right]^2}\,\Im\!\left[\frac{-1}{\epsilon(\bq,\omega)}\right],
 \label{Prob}
\end{eqnarray}
where we have set the reflection coefficient to unity.

\end{section}

\begin{section}{Comparison with semi-classical model}

In this section, we present the stopping and image forces calculated with dielectric functions from the SCKE model
\cite{Radovic_2008} and the MDF-RPA model \cite{Wunsch_2006,Hwang_2007} for 
free graphene ($\ebg^0=1$) and a vanishing damping rate ($\gamma\rightarrow0$). The results for both forces are normalized by
$F_0=Z^2e^2/(4z_0^2)$, the magnitude of the classical image force on a static point charge a distance $z_0$ from a perfect
conductor, to better reveal differences between the two models.

Before proceeding, we note that the infinite upper limits of the $q$ integrals in Eqs.\ (\ref{stopping2}) and (\ref{image2})
cause both forces to diverge as the distance $z_0$ goes to zero. This behaviour, which also occurs in models of solid surfaces,
\cite{Barton_1979,Lucas_1979} should not be a major concern because the restriction $z_0>a$ is necessary to ensure the validity
of the MDF approximation. However, if one would like to extend the results for the stopping and image forces to include small
distances, a standard procedure to eliminate the divergence at $z_0$ is to adopt a high-momentum cut-off.
\cite{Barton_1979,Lucas_1979} For studies of electronic processes in graphene with no external charges, it is common to impose a
sharp cut-off at approximately $k_c$. \cite{Wunsch_2006,Barlas_2007,Castro_2009} However, there are other mathematical methods
for imposing a cut-off besides this sharp truncation of the $q$ integration. \cite{Lucas_1979} As discussed in
Ref.\cite{Lucas_1979}, the use of an exponential cut-off function $\mathrm{e}^{-q/k_c}$ in Eqs.\ (\ref{stopping2}) and
(\ref{image2}) would simply amount to a shift of the $z_0$ coordinate by a distance on the order of the lattice constant, which
is sometimes referred to as an effective image plane. \cite{Lucas_1979,Gumbs_2009} We therefore evaluate the stopping and image
forces using Eqs.\ (\ref{stopping2}) and (\ref{image2}) with infinite upper limits in the $q$ integrals, and if one would like
an estimate of the order of magnitude of these forces at small distances we note that a suitable shift of the $z_0$ axis may be
chosen. It should also be mentioned that in an RPA model that takes into account the finite size of graphene's $\pi$ electron
orbitals, the divergence of these forces as $z_0\rightarrow 0$ can be removed by the resulting structure factor, which provides
a physically motivated algebraic cut-off function. \cite{Shung_1986,Lucas_1979}

\begin{figure}
\centering
\includegraphics[width=0.5\textwidth]{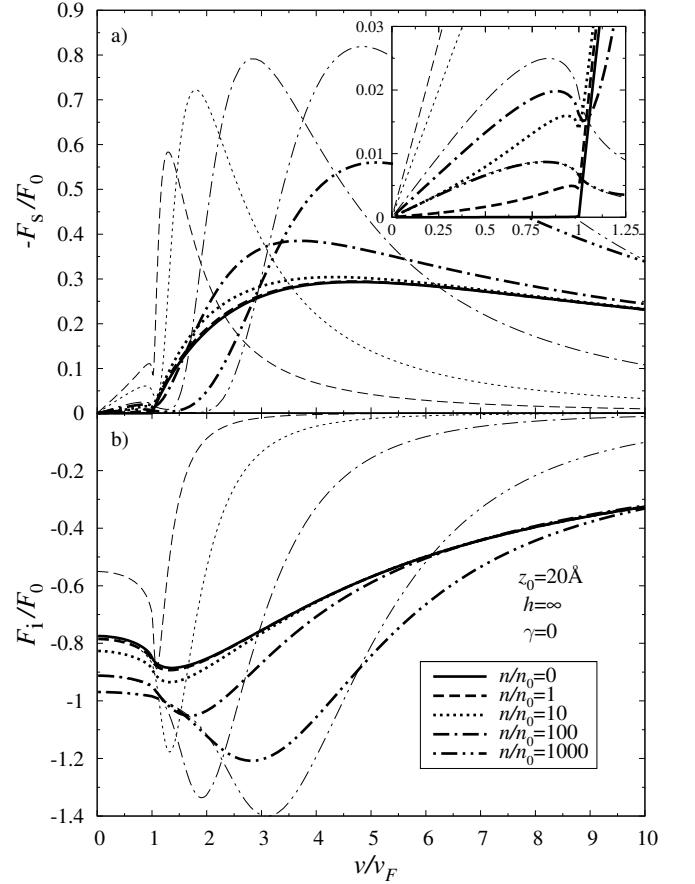}
\caption{ The stopping force (a) and image force (b) normalized by $F_0=Z^2e^2/(4z_0^2)$ and shown as a function of the reduced
speed $v/v_F$ of a proton ($Z=1$) moving at a distance $z_0$ = 20 \AA\ above free graphene ($h\rightarrow\infty$) for several
values of the reduced charge-carrier density $n/n_0$, where $n_0=10^{11}$ cm$^{-2}$. The thick and thin lines represent the
results from the MDF-RPA and SCKE models with vanishing damping ($\gamma=0$), respectively.}
\end{figure}

In Fig.\ 1, we compare the velocity dependence of the normalized stopping and image forces on a proton ($Z=1$) moving at a
distance $z_0=20$ \AA\ above free graphene in the MDF-RPA model (thick lines) and in the SCKE model (thin lines) for a broad
range of densities. For intrinsic graphene ($n=0$), note that both forces vanish in the SCKE model but they arise from
inter-band SPEs in the MDF-RPA model. Furthermore, it can be seen that the results from the SCKE model agree with those from the
MDF-RPA model only for high densities, and that this agreement is better for low particle speeds ($v<v_F$) than for high
particle speeds ($v>v_F$). The large difference between the MDF-RPA and SCKE models at high particle speeds is due to the
presence of a plasmon line given by $\omega=\omega_p(q)$, \cite{Vafek_2006,Ryzhii_2007,Radovic_2008} where
$\omega_p(q)=v_F(q+q_s)\sqrt{q/(q+2q_s)}>qv_F$ and $q_s\equiv 4r_sk_F$ is the Thomas-Fermi (TF) inverse screening length.
\cite{Ando_2006,Hwang_2007,Radovic_2008} Specifically, the energy loss for high particle speeds in the SCKE model is dominated
by the undamped plasmon at frequency $\omega_p(q)$, while the presence of the inter-band SPE continuum in the MDF-RPA model for
$\omega/v_F>\text{max}\left(q,2k_F-q\right)$ causes a strong Landau damping of the plasmon, \cite{Wunsch_2006,Hwang_2007}
thereby producing much weaker velocity dependencies. However, even for these high particle speeds it appears that the SCKE model
may be partially applicable under the condition $z_0k_F \gg 1$, which requires heavy doping of graphene in order to reduce the
significance of the inter-band SPEs.

\begin{figure}
\centering
\includegraphics[width=0.5\textwidth]{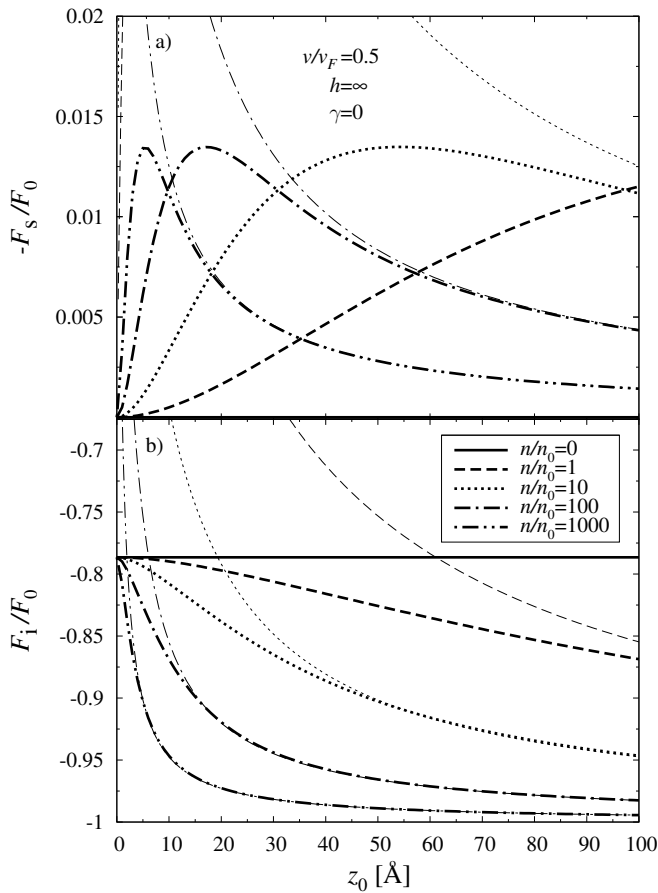}
\caption{ The stopping force (a) and image force (b) normalized by $F_0=Z^2e^2/(4z_0^2)$ and shown as a function of the distance
$z_0$ of a proton ($Z=1$) moving at a reduced speed $v/v_F = 0.5$ above free graphene ($h\rightarrow\infty$) for several values
of the reduced charge-carrier density $n/n_0$, where $n_0=10^{11}$ cm$^{-2}$. The thick and thin lines represent the results
from the MDF-RPA and SCKE models with vanishing damping ($\gamma=0$), respectively.}
\end{figure}

The difference between the SCKE and MDF-RPA models at low particle speeds is further analyzed in Fig.\ 2. Using the same set of
densities as in Fig.\ 1, we compare the reduced stopping and image forces on a proton ($Z=1$) moving at a speed $v=v_F/2$ above
free graphene as a function of the particle distance, $z_0$. At such low speeds, one can see that the agreement between the SCKE
and MDF-RPA models is better for the image force than it is for the stopping force. It follows from Fig.\ 2 that the condition
$z_0k_F > 1$ may suffice as a rough criterion for the application of the SCKE model at low speeds ($v<v_F$). This condition is
far less restrictive than $z_0k_F \gg 1$, which is required for the application of the SCKE model at high speeds ($v>v_F$). We
therefore discontinue further analysis of the SCKE model at high speeds, and turn our focus to analyzing various parameters in
the MDF-RPA model alone.

\end{section}

\begin{section}{MDF-RPA with vanishing damping}

In this section, we use the MDF-RPA dielectric function with a vanishing damping rate ($\gamma \rightarrow 0$) to evaluate the
stopping and image forces. We first investigate the effects of a finite graphene-substrate gap, and then discuss two important
cases with a zero gap: intrinsic graphene ($n=0$) and vanishing particle speeds ($v \rightarrow 0$).

\subsection{Effects of a finite gap}

\begin{figure}
\centering
\includegraphics[width=0.5\textwidth]{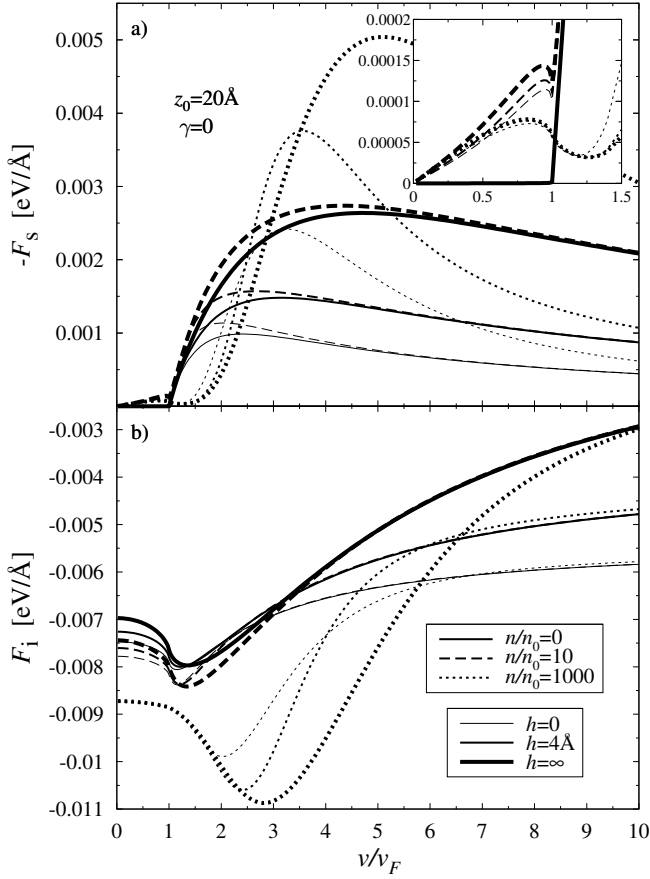}
\caption{ The stopping force (a) and image force (b) from the MDF-RPA model with vanishing damping ($\gamma=0$)
shown as a function of the reduced speed $v/v_F$ of a proton ($Z=1$) moving at a distance $z_0$ = 20 \AA\ above graphene on a
SiO$_2$ substrate ($\es\approx3.9$). Results are shown for several values of the gap height $h$ and several values of the
reduced charge-carrier density $n/n_0$, where $n_0=10^{11}$ cm$^{-2}$.}
\end{figure}

We now assume that graphene is supported by a SiO$_2$ substrate ($\es\approx 3.9$) and explore the effects of a variable gap
height. It should be noted that the mean gap height between graphene and a SiO$_2$ substrate has been measured as 4.2 \AA,
 \cite{Ishigami_2007} which is comparable to the equilibrium distance of 3.6 \AA\ found in \emph{ab inito} calculations between
graphene and the topmost atomic plane of a SiO$_2$ substrate. \cite{Romero_2008} However, we note that $h$ is defined in
Ref.\cite{Radovic_2008} as the position of an effective substrate surface plane where boundary conditions on the electrostatic
fields are imposed. Thus, while the measured and theoretically obtained values of the graphene-substrate gap serve as a guide
for the value of $h$, there is an uncertainty in $h$ on the same order as the shift of the $z$ axis discussed earlier. In this
subsection, we consider the gap heights $h\rightarrow\infty$ for free graphene, $h$ = 4 \AA\ for a realistic value, and $h=0$
for the zero gap commonly considered in the literature.

In Fig.\ 3, we compare the velocity dependence of the stopping and image forces on a proton moving at a distance $z_0=20$ \AA\
above graphene for several gap heights and densities. For low particle speeds ($v<v_F$), the gap height has a relatively small
influence on the stopping and image forces that diminishes as the charge-carrier density increases and effectively screens out
the graphene-substrate gap. The density, $n$, is therefore the most important parameter in the low-speed behaviour of both
forces. For higher particle speeds, the charge carriers in graphene are not as effective in screening out the graphene-substrate
gap, and hence the gap height has a much stronger effect on the stopping and image forces. In particular, Fig.\ 3 shows that for
sufficiently high speeds ($v \gg v_F$) an increase in the gap height tends to increase the strength of the stopping force and
decrease the strength of the image force, but there is a range of moderate speeds for which this trend is reversed. Also note
that for sufficiently high speeds, all MDF-RPA stopping and image forces approach the intrinsic case, $n=0$.

\begin{figure}
\centering
\includegraphics[width=0.5\textwidth]{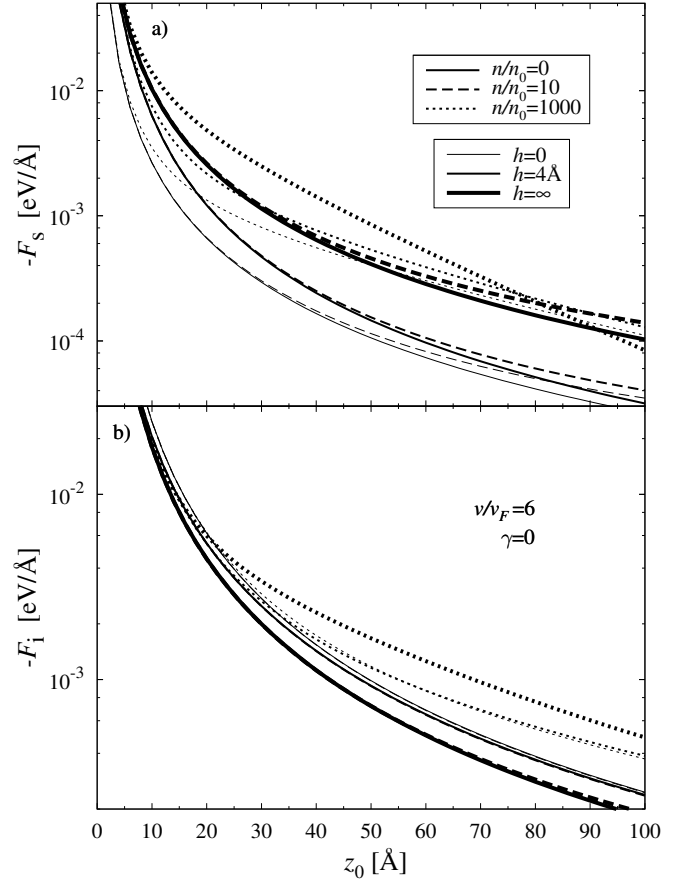}
\caption{ The stopping force (a) and image force (b) from the MDF-RPA model with vanishing damping ($\gamma=0$) shown as a
function of the distance $z_0$ of a proton ($Z=1$) moving at a reduced speed $v/v_F=6$ above graphene on a SiO$_2$ substrate
($\es\approx3.9$). Results are shown for several values of the gap height $h$ and several values of the reduced charge-carrier
density $n/n_0$, where $n_0=10^{11}$ cm$^{-2}$.}
\end{figure}

The effect of the gap height at high speeds is further explored in Fig.\ 4, which shows the distance dependence of the stopping
and image forces on a proton moving at a moderately high speed $v=6v_F$ above graphene for the same gap heights and densities as
in Fig.\ 3. In Fig.\ 4, it can be seen that the gap height has a strong influence on both forces for all distances. One may
conclude that in the MDF-RPA model, as in the SCKE model, \cite{Radovic_2008} any uncertainty or local variations in the gap
height across graphene can lead to large fluctuations in the stopping and image forces, particularly for high particle speeds.

\subsection{Intrinsic graphene with a zero gap}

We now take advantage of the simplicity of the MDF-RPA dielectric function for intrinsic graphene with a vanishing damping rate
\cite{Wunsch_2006,Hwang_2007} to evaluate the stopping and image forces analytically for a zero gap. In this case, it is worth
noting that the distance dependence of both forces can be factored out as $F_0=Z^2e^2/(4z_0^2)$. For the stopping force, we find
\begin{equation}
\Fs^0=-\frac{F_0}{\ebg^0}\, \frac{\rho_s}{\nu}\left[ 1-\left( 1+\frac{\nu^2-1}{\rho_s^2} \right)^{-1/2} \right],
\end{equation}
where $\rho_s\equiv\pi r_s/2$ and $\nu\equiv v/v_F$. As seen from the thick solid curve in Fig.\ 1(a), this expression is
subject to the velocity threshold constraint $v>v_F$, which is a consequence of the inter-band SPEs yielding the loss function
$-\Im\epsilon(q,\omega)>0$ only for $\omega>qv_F$. \cite{Wunsch_2006,Hwang_2007} For sufficiently high particle speeds ($v\gg
v_F$), one obtains an asymptotic form $\Fs^0\sim -(\pi/8)\,\left[Ze^2/(\ebg^0 z_0)\right]^2/\left(\hbar v\right)$ that is
independent of $v_F$. It is interesting to note that the MDF-RPA stopping forces for all densities approach this high-speed
asymptotic limit, as seen in Fig.\ 1(a) for free graphene and in Fig.\ 3(a) for various gap heights.

The corresponding expression for the image force on a charged particle moving above intrinsic graphene, $\Fi^0$, is rather
cumbersome. We therefore define the \emph{effective} background dielectric constant $\ebg^*$ by writing the image force in the
form $\Fi^0=F_0\left(1/\ebg^*-1\right)$, and in Fig.\ 5 we present $\ebg^*$ as a function of the particle speed $v$ and the
\emph{actual} background dielectric constant $\ebg^0=(\es+1)/2$. We do this for background dielectric constants ranging from
free graphene ($\ebg^0=1$) to a HfO$_2$ substrate ($\ebg^0\approx 14)$. For vanishing particle speeds ($v\rightarrow0$), one
finds $\ebg^*\rightarrow\ebg^0\left(1+\rho_s\right)\approx \ebg^0+\frac{\pi}{2}\frac{v_B}{v_F}$, which is a well-known result
for the contribution of inter-band SPEs to the static limit of the MDF-RPA dielectric constant for intrinsic graphene.
\cite{Ando_2006,Hwang_2007} For sufficiently fast particles ($v\gg v_F$), graphene becomes ``transparent'' and one recovers the
case of a bare substrate, $\ebg^*\rightarrow\ebg^0$, with an accuracy to the order of $\ebg^*-\ebg^0\sim v^{-1}$. Again, it is
interesting to note that the MDF-RPA image forces for all densities eventually approach this high-speed asymptotic limit for
intrinsic graphene, as seen in Figs.\ 1(b) and 3(b).

\begin{figure}
\includegraphics[width=0.5\textwidth]{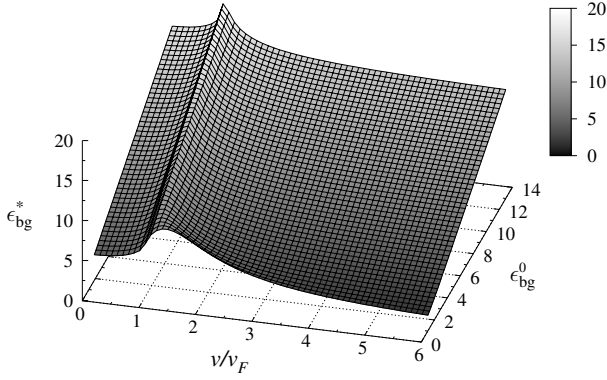}
\caption{The reduced image force $\Fi^0$ from the MDF-RPA model with vanishing damping ($\gamma=0$) for a proton ($Z=1$) moving
over intrinsic graphene ($n=0$) on a substrate with dielectric constant $\es$ and a zero gap ($h=0$). The force is expressed as
an effective background dielectric constant $\ebg^*=1/\left[1+(4z_0^2/e^2)\,\Fi^0\right]$, and is shown as a function of the
reduced proton speed $v/v_F$ and the actual background dielectric constant $\ebg^0=(\es+1)/2$.}
\end{figure}

\subsection{Vanishing particle speed with a zero gap}

In this subsection, we consider the density dependence of the stopping and image forces from the MDF-RPA model in the limit of
vanishing particle speed ($v\rightarrow 0$) for a zero gap. For sufficiently small speeds, the inset in Fig.\ 1(a) shows that
the stopping force is proportional to the particle speed $v$, and hence we can define a friction coefficient $\eta$ through the
equation $\Fs=-\eta v$. To evaluate $\eta$, it is sufficient to expand the loss function
$\Im\!\left[-1/\epsilon(q,\omega)\right]$ to the first order in $\omega$ and use the resulting expression in Eq.\
(\ref{stopping2}) to find the stopping force. \cite{Alducin_2007} We find the expression for the continuum of low-energy,
intra-band SPEs, subject to the constraint $q\le 2k_F$, to be
\begin{eqnarray}
-\Im\epsilon^{-1}(q,\omega)\approx\frac{2r_s}{\ebg^0\etf^2(q)}\,\frac{\omega}{qv_F}\sqrt{\left(\frac{2k_F}{q}\right)^2-1},
\label{loss}
\end{eqnarray}
where $\etf(q)=1+q_s/q$ is the TF dielectric function. \cite{Ando_2006,Hwang_2007,Radovic_2008} The friction coefficient is then
given by $\eta=2\pi\hbar\, n Z^2 I(4z_0k_F,r_s)$, where the function $I(a,r_s)$ is defined in Eq.\ (4) of Ref.\cite{Adam_2008}

The case $z_0k_F\ll 1$ is particularly interesting because, unlike in the SCKE model, the constraint $q\le 2k_F$ in Eq.\
(\ref{loss}) causes $I(4z_0k_F,r_s)$ to remain finite even as $z_0k_F\rightarrow 0$. \cite{Radovic_2008} The friction
coefficient for a charge moving very close to graphene in the MDF-RPA model is therefore given by $\eta=2\pi\hbar\, n Z^2
I(0,r_s)$, and is proportional to the charge-carrier density $n$. In the opposite case, $z_0k_F\gg 1$, one recovers the TF
result for the friction coefficient given in Eq.\ (37) of Ref.\cite{Radovic_2008}, which yields an asymptotic form
$\eta=Z^2\hbar/(32z_0^3\sqrt{\pi n})$ that is independent of $v_F$. These two limiting cases for the friction coefficient can be
observed in Fig.\ 2(a), which shows the MDF-RPA stopping forces for a particle speed $v=v_F/2$ that is very near the static
limit. Recalling that the stopping force $F_s\approx-\frac{1}{2}v_F\eta$ is normalized by $F_0=Z^2e^2/(4z_0^2)$ in Fig.\ 2(a),
note that all MDF-RPA curves are proportional to $z_0^2$ for short distances and fall off as $z_0^{-1}$ for large distances. The
transition between these two behaviours occurs around the peaks of the curves at $z_0k_F\sim 1$.

In the limit of vanishing particle speed, the image force is closely related to the well-studied problem of static screening of
an external charge, for which the dielectric function in Eq.\ (\ref{eps}) reduces to $\epsilon(q,0)\equiv\ebg^0+\frac{2\pi
e^2}{q}\Pi_s(q)$, where $\Pi_s(q)$ is the static MDF-RPA polarization function for graphene given in Appendix A.
\cite{Ando_2006,Sarma_2007,Wunsch_2006,Hwang_2007} Although the $\omega$ integration in Eq.\ (\ref{image2}) is trivial with this
dielectric function, the remaining $q$ integration cannot be completed analytically. However, expressions for the image force in
the two limiting cases of $z_0k_F$ can be given explicitly. For $z_0k_F\ll 1$, the MDF-RPA image force reduces to the static
limit for intrinsic graphene, $\Fi^0=F_0\!\left[1/\!\left(\ebg^0+\frac{\pi}{2}\frac{v_B}{v_F}\right)-1\right]$. In the opposite
case, $z_0k_F\gg 1$, one easily recovers the TF result for the static image force given in Eq.\ (38) of Ref.\cite{Radovic_2008},
which yields asymptotically $\Fi\sim-F_0$. Fig.\ 2(b) clearly shows the transition between these two cases for the particle
speed $v=v_F/2$, which is a good approximation to the static limit. Note that all normalized MDF-RPA image forces fall between
the limits $\left(1+\frac{\pi}{2}\frac{v_B}{v_F}\right)^{-1}-1\approx-0.78$ and $-1$ in Fig.\ 2(a), with a broad transition
occurring at $z_0k_F\sim1$.

\end{section}

\begin{section}{MDF-RPA with finite damping}

In this section, we use the MDF-RPA model with a finite damping rate $\gamma>0$ (by means of the Mermin dielectric function) to
evaluate the stopping and image forces. Since the exact value of the damping rate is not known, we first treat it as a fitting
parameter and come up with an estimate for $\gamma$ by comparing the MDF-RPA model with a finite damping rate to experimental
data for the HREELS spectra of graphene on a SiC substrate. We then use this estimate to investigate the effects of the damping
rate on the stopping and image forces for free graphene ($h\rightarrow\infty$), with a special focus on its role in the friction
of low-speed particles.

It should be mentioned that introducing a finite $\gamma$ into the MDF-RPA dielectric function for graphene to create the Mermin
dielectric function, $\epsilon_M(q,\omega,\gamma)$, is a non-trivial matter, as described in Appendix A. There is, however, a
significant advantage in using such a dielectric function in numerical calculations of the stopping and image forces.
Specifically, expressions for the MDF-RPA dielectric function in the strict $\gamma=0$ case make numerical integrations of Eqs.\
(\ref{stopping2}) and (\ref{image2}) difficult as the boundary of the integration domain, $0\le\omega\le qv$, intersects the
delta-like plasmon line. A small but finite $\gamma$, however, broadens the plasmon line to allow for a simple numerical
treatment of this delta-like behaviour. Throughout this paper, we have therefore reproduced the strict $\gamma=0$ limit of the
MDF-RPA dielectric function \cite{Wunsch_2006,Hwang_2007} by using the value $\gamma=10^{-3}\times v_Fk_0$, where $k_0=\sqrt{\pi
n_0}$ and $n_0=10^{11}$ cm$^{-2}$.

\subsection{Comparison with HREELS experiment}

To obtain a reasonable estimate for $\gamma$, we compare the MDF-RPA model with finite damping to the experimental data of Liu
\emph{et al}.\ \cite{Liu_2008} for the HREELS spectra of graphene on a SiC substrate. Since the focus of this paper is on
slow-moving particles that are unable to excite graphene's $\sigma$ electrons, this HREELS experiment is more relevant to the
present work than the EELS experiments on graphene and other carbon nanostructures.
\cite{Stephan_2002,Kramberger_2008,Eberlein_2008} We note, however, that the effects of a SiC substrate on graphene are not as
well understood as those of a SiO$_2$ substrate. Without entering the current debate, \cite{Trevisanutto_2008,Zhou_2007} we
simply neglect any changes in graphene's $\pi$-band structure that may result from a hybridization of its $\pi$ orbitals with
the substrate and treat the gap height, $h$, as a free parameter.

\begin{figure}
\centering
\includegraphics[width=0.5\textwidth]{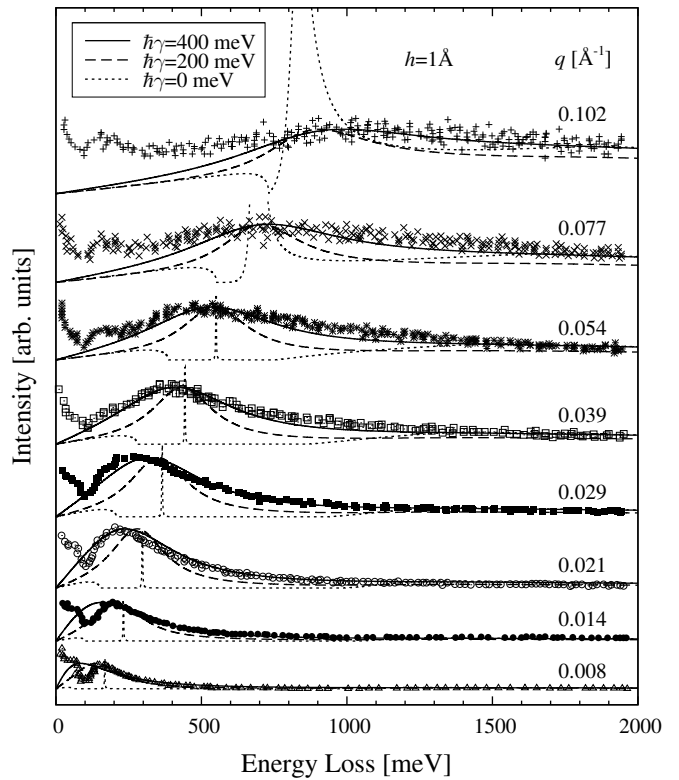}
\caption{The Mermin loss function (in arbitrary units) versus the energy loss for graphene with a charge-carrier density
$n=2\times 10^{13}$ cm$^{-2}$) supported on a SiC substrate with static dielectric constant $\es=9.7$ and a gap height $h$ = 1
\AA. Model results are shown for damping rates $\hbar\gamma$ = 0, 200, and 400 meV, while symbols show the HREELS experimental
data from Ref.\cite{Liu_2008}}
\end{figure}

Since the measurements in Ref.\cite{Liu_2008} indicate that the maximum HREELS yields occur in the direction of near-specular
reflection of slow incident electrons, it is appropriate to use Eq.\ (\ref{Prob}) for the probability density of exciting the
mode $(q,\omega)$. However, since we are not aware at this time of specific details of the experimental procedure that
predominantly affect the low frequency range of the HREELS spectra via the prefactor in Eq.\ (\ref{Prob}),
\cite{Liu_2008,Ibach_1982} we simply focus on the Mermin loss function $\Im\!\left[-1/\epsilon_M(q,\omega,\gamma)\right]$ and
assume that it gives the correct order of magnitude for spectral widths outside of this low frequency range. Therefore, in Fig.\
6 we display a tentative comparison between the HREELS data \cite{Liu_2008} and the Mermin loss function with $\hbar\gamma=0$,
200, and 400 meV and a gap height of 1~\AA\ for wavenumbers ranging from 0.008 to 0.102 \AA$^{-1}$. The SiC subtrate is treated
in the static mode with dielectric constant $\es=9.7$, and the equilibrium density in graphene is set at $n=2\times 10^{13}$
cm$^{-2}$ (hence $\varepsilon_F\approx 570$ meV and $k_F\approx$ 0.08 \AA$^{-1}$) to match experimental conditions. Note that
since the HREELS data is scaled arbitrarily, the Mermin loss functions for $\hbar\gamma=200$ and 400 meV are scaled so that the
maximum peak heights coincide with those from the experiment. For $\hbar\gamma=0$, however, the singular plasmon peak prevents
such a scaling, and so the Mermin loss function is scaled by the same factor as for the $\hbar\gamma=400$ meV loss function.

In Fig.\ 6, the range $q\lesssim k_F$ is particularly interesting because the Mermin loss function with $\hbar\gamma=0$ exhibits
three distinct features for these wavenumbers: a continuous spectrum of intra-band SPEs for $0<\omega<v_Fq$, a continuous
spectrum of inter-band SPEs for $\omega>v_F(2k_F-q)$, and a narrow plasmon line at $\omega=\omega_p(q)$ in the otherwise void
interval $v_Fq<\omega<v_F(2k_F-q)$. The fact that these three features are not visible in the experimental HREELS spectra can be
tentatively explained by assuming that a large enough damping rate $\gamma$ exists, due to various scattering mechanisms, that a
broadened plasmon line merges into the two regions of SPEs to form a single peak that follows approximately the original plasmon
dispersion curve, $\omega=\omega_p(q)$. Note that a broadening of the plasmon line for the $\hbar\gamma=0$ loss function does
occur for the wavenumbers $q=0.077$ and 0.102 \AA$^{-1}$ as the plasmon line crosses the boundary $\omega=v_F(2k_F-q)$ and
enters into the region of inter-band SPEs, in which collective plasma oscillations decay into SPEs in a way that can be
described by a finite Landau damping rate, $\gamma_L$. \cite{Pines_1962} For the Mermin loss function with a phenomenological
damping rate $\gamma$, however, a broadening of the plasmon line occurs for all $q$. Given that we have neglected the effects of
temperature, the kinematic prefactor, and the reflection coefficient in Eq.\ (\ref{Prob}), which all give rise to the
low-frequency features in the HREELS spectra, Fig.\ 6 shows that a reasonably good qualitative agreement with the experiment can
be achieved by using a gap height $h=1$~\AA\ and a damping rate $\hbar\gamma=400$ meV.

\begin{figure}
\centering
\includegraphics[width=0.5\textwidth]{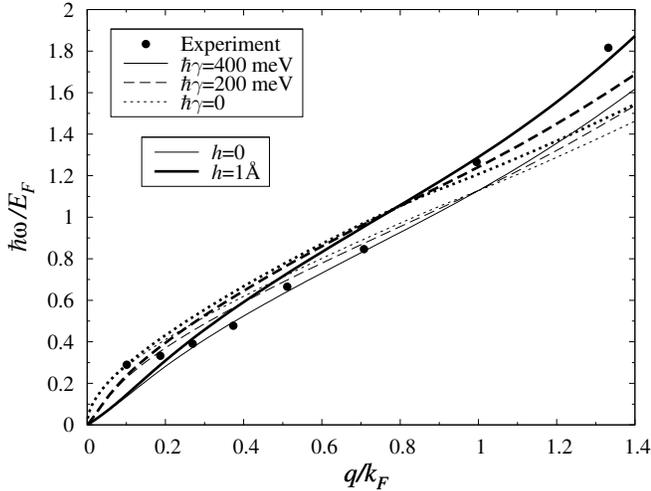}
\caption{The peak positions of the Mermin loss function shown as a function of the wavenumber $q/k_F$ for graphene with a
charge-carrier density $n=2\times 10^{13}$ cm$^{-2}$ supported on a SiC substrate with static dielectric constant $\es=9.7$.
Model results are shown for damping rates $\hbar\gamma$ = 0, 200, and 400 meV, as well as for gap heights $h=0$ and 1 \AA.
Filled circles show the HREELS experimental data from Ref.\cite{Liu_2008}}
\end{figure}

In Fig.\ 7, we compare the peak positions of the model spectra in Fig.\ 6 with the experimental HREELS data. \cite{Liu_2008} In
addition to the best fit gap height $h=1$~\AA, we also include the zero gap case to demonstrate the effect of the gap height.
Note, however, that this effect is limited to shifting the spectral peak positions, and therefore does not significantly affect
the estimate of the damping rate. In the long wavelength limit ($q\rightarrow0$), it can be seen that the plasmon dispersion
curve $\omega=\omega_p(q)$ for a vanishing damping rate exhibits the $\sqrt{q}$ behaviour of a classical 2D electron gas,
\cite{Wunsch_2006,Hwang_2007} but as the damping rate increases the plasmon dispersion curve falls off in a more quasi-acoustic
manner. It is worth noting, however, that none of the curves are able to describe the trend of the experimental data at long
wavelengths. From Fig.\ 6, it can be seen that there is a strong coupling of graphene's $\pi$ electron excitations with the
non-dispersing Fuchs-Kliever surface phonon mode in the SiC substrate at the frequency $\approx$ 116 meV. \cite{Liu_2008,
Nienhaus_1995} We note that a dynamic treatment of the substrate phonon excitation through Eq.\ (\ref{eq:esub}), the use of a
$q$-dependent damping rate, and the inclusion of low-frequency features through the prefactor in Eq.\ (\ref{Prob}) may be
necessary for accurate modeling of the experimental HREELS spectra in the full range of frequencies. However, further discussion
on these points would go beyond the scope of the paper, and we merely conclude that a reasonable value for the damping rate
$\hbar\gamma$ is on the order of several hundred meV.

\subsection{Results for MDF-RPA with finite damping}

In this subsection, we discuss the effects of a finite damping rate on the stopping and image forces calculated with the Mermin
dielectric function for free graphene with several charge-carrier densities. Since it was found that a reasonable value for
$\hbar\gamma$ is on the order of several hundred meV, we again consider the values $\hbar\gamma=0$, 200, and 400 meV, or
equivalently $\gamma/(v_Fk_0)\approx0$, 5, and 10, where $k_0=\sqrt{\pi n_0}$ and $n_0=10^{11}$ cm$^{-2}$.

\begin{figure}
\centering
\includegraphics[width=0.5\textwidth]{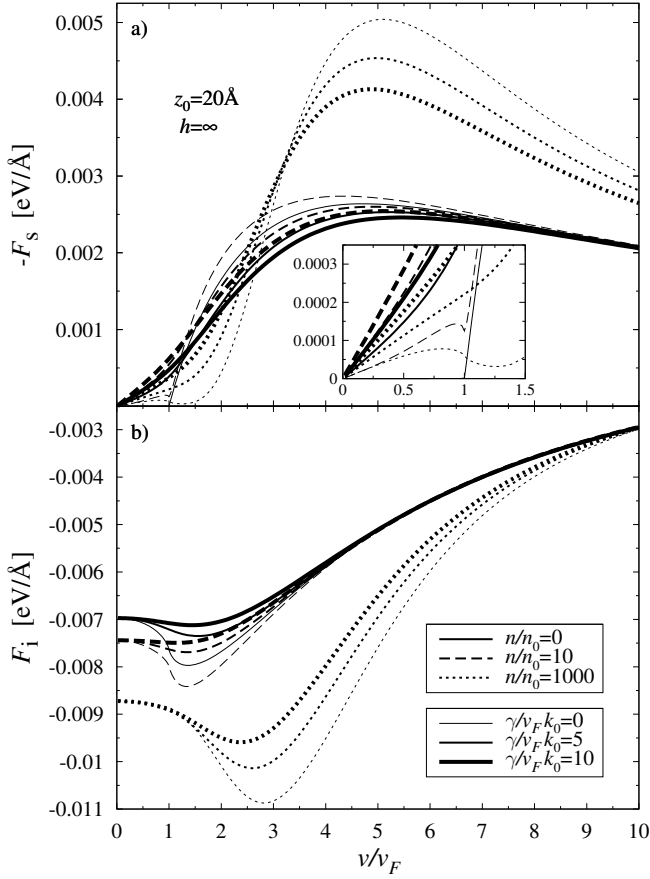}
\caption{ The stopping force (a) and image force (b) from the MDF-RPA model with a finite damping rate shown as a function of
the reduced speed $v/v_F$ of a proton ($Z=1$) moving at a distance $z_0$ = 20 \AA\ above free graphene ($h\rightarrow\infty$).
Results are shown for several values of the reduced damping rate $\gamma/(v_Fk_0)$, where $k_0=\sqrt{\pi n_0}$, and for several
values of the reduced charge-carrier density $n/n_0$, where $n_0=10^{11}$ cm$^{-2}$.}
\end{figure}

In Fig.\ 8, we examine the velocity dependence of the stopping and image forces for the various damping rates and densities. For
medium to high speeds ($v>v_F$), it can be seen that the strength of both forces decreases significantly as the damping rate
increases, and that this effect diminishes faster for the image force than for the stopping force as the particle speed
increases. For low speeds ($v<v_F$), the fact that the image force is unaffected by the damping rate can be understood through
Eq.\ (\ref{Pi_M}), from which it can be seen that the static limit of the Mermin polarization reduces to the static limit of the
MDF-RPA polarization, $\Pi_M(q,0,\gamma)=\Pi_s(q)$.

Fig.\ 8(a) also shows an interesting relationship between the damping rate, $\gamma$, and the density, $n$, at low speeds. One
can see that the strength of the stopping force tends to increase with $\gamma$ at a fixed density, but at a fixed $\gamma$ the
stopping force tends to peak at some intermediate density. This trend is further confirmed in Fig.\ 9(a), which shows the
stopping force as a function of the particle distance $z_0$ for low particle speeds. For high particle speeds, however, Fig.\
9(b) shows that the distance dependence of the stopping force is affected more by the density than the damping rate.

\begin{figure}
\centering
\includegraphics[width=0.5\textwidth]{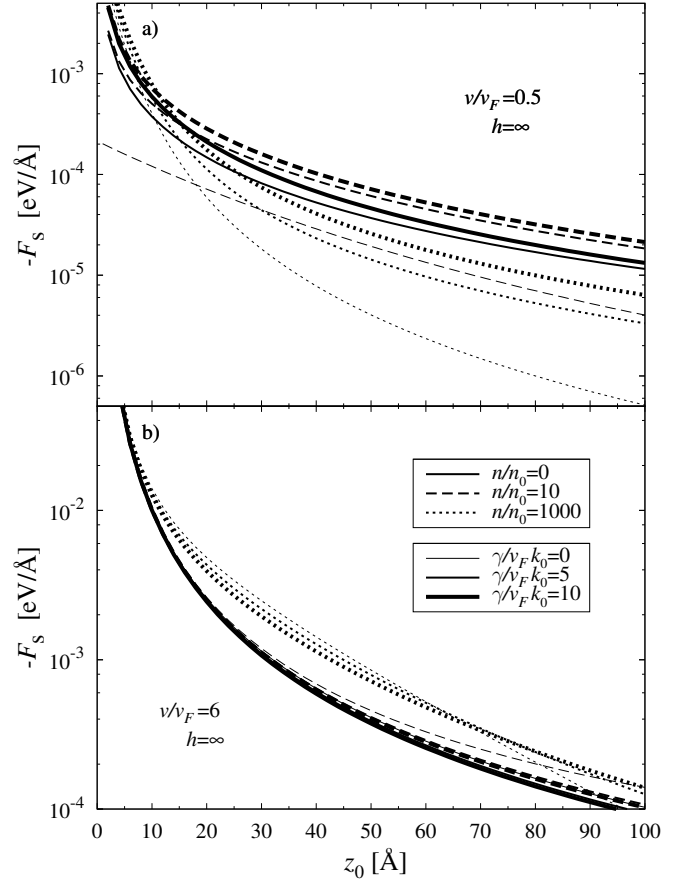}
\caption{ The stopping force from the MDF-RPA model with a finite damping rate shown as a function of the distance $z_0$ of a
proton ($Z=1$) moving at the low speed $v=v_F/2$ (a) and the moderately high speed $v=6v_F$ (b) above free graphene
($h\rightarrow\infty$). Results are shown for several values of the reduced damping rate $\gamma/(v_Fk_0)$, where $k_0=\sqrt{\pi
n_0}$, and for several values of the reduced charge-carrier density $n/n_0$, where $n_0=10^{11}$ cm$^{-2}$.}
\end{figure}

\subsection{Friction coefficient}

The stopping forces in the inset of Fig.\ 8(a) suggest that, as for the case of vanishing damping, the concept of friction may
be useful in the case of finite damping for low particle speeds. Again, we define the friction coefficient $\eta$ through the
equation $\Fs=-\eta v$. Although we wish to focus on the case of free graphene ($h\rightarrow\infty$), we provide analytic
results for the slightly more general case of a zero gap and note that the case of free graphene can be recovered by taking
$\es=1$, and hence $\ebg^0=1$. To evaluate $\eta$, we expand the Mermin loss function for a zero gap to the first order in
$\omega$, which gives
\begin{eqnarray}
\label{eq:mermin_w} -\Im\epsilon_M^{-1}(q,\omega,\gamma)\approx
\frac{\omega}{\gamma} \frac{2\pi e^2}{q}
 \frac{\Pi_s(q)}{\left[\ebg^0+\frac{2\pi
e^2}{q}\Pi_s(q)\right]^2}\left[-1+
\frac{\Pi_s(q)}{\Pi(q,i\gamma)}\right],
\end{eqnarray}
Eq.\ (\ref{eq:mermin_w}) can then be substituted into the stopping force, Eq.\ (\ref{stopping2}), to get an expression for the
friction coefficient. Since the resulting integral cannot be evaluated analytically for an arbitrary density $n$, we first
consider the case of intrinsic graphene ($n=0$), for which
\begin{eqnarray}
\Pi(q,i\gamma)=\frac{1}{4\hbar}\frac{q^2}{\sqrt{\gamma^2+\left(qv_F\right)^2}}, \label{Pi_gamma}
\end{eqnarray}
and hence $\Pi_s(q)=q/(4\hbar v_F)$. Using these expressions in Eq.\ (\ref{eq:mermin_w}), the friction coefficient for intrinsic
graphene is given by
\begin{eqnarray}
\eta_0=\frac{\pi
Z^2e^2}{4\zeta_0v_F}\left(\frac{\gamma}{v_F}\right)^2\frac{\frac{\pi}{2}\frac{v_B}{v_F}}{\left(\ebg^0+
\frac{\pi}{2}\frac{v_B}{v_F}\right)^2} \times \nonumber \\
\left[-\frac{1}{2\pi\zeta_0}+Y_0(\zeta_0)-Y_1(\zeta_0)-H_0(\zeta_0)+H_1(\zeta_0)\right],
\label{eta_zero}
\end{eqnarray}
where $\zeta_0\equiv 2z_0\gamma/v_F$, and $Y_\nu$ and $H_\nu$ are the Bessel function of the second kind and the Struve
function, respectively. By combining the leading terms of the series expansions for large and small $\zeta_0$ in this expression
for $\eta_0$, one obtains the simple but surprisingly accurate formula (with a maximum error of approximately 3\% at
$\zeta_0\approx 1$)
\begin{eqnarray}
\eta_0\approx\frac{Z^2e^2}{8v_F}\frac{\frac{\pi}{2}\frac{v_B}{v_F}}{\left(\ebg^0+\frac{\pi}{2}\frac{v_B}{v_F}\right)^2}
\frac{1}{z_0\!\left(z_0+\frac{v_F}{\gamma}\right)}. \label{approx}
\end{eqnarray}

We now consider the friction coefficient $\eta$ for an arbitrary density $n$. In Fig.\ 10, we display the normalized friction
coefficient $\eta/\eta_0$ for free graphene ($\ebg=1$) as a function of the reduced damping rate $\gamma/(v_Fk_0)$, where
$k_0=\sqrt{\pi n_0}$, and the reduced charge-carrier density $n/n_0$, where $n_0=10^{11}$ cm$^{-2}$. It can be seen that the
friction coefficient for a slow particle moving at a distance $z_0$ = 20 \AA\ above free graphene is a rather complicated
function of $\gamma/(v_Fk_0)$ and $n/n_0$, but the qualitative behaviour of this function can be understood. Recall from
subsection III.C that the friction coefficient for a vanishing damping rate goes as $\eta\propto n$ for $z_0k_F \ll 1$ and as
$\eta\propto 1/(z_0^3n)$ for $z_0k_F \gg 1$, where $k_F=\sqrt{\pi n}$. For the distance $z_0=20$ \AA, the transition between
these two limiting behaviours occurs at $n/n_0\approx 80$. For a finite damping rate, we find in Eq.\ (\ref{approx}) that the
friction coefficient goes as $\eta\propto \gamma/(z_0v_F)$ for $z_0\gamma/v_F\ll 1$ and as $\eta\propto z_0^{-2}$ for
$z_0\gamma/v_F\gg 1$. For the distance $z_0$ = 20 \AA, the transition between these two limiting behaviours occurs at
$\gamma/(v_Fk_0)\approx 9$. Note how a saddle point develops at these two values for the density and damping rate in the surface
plotted in Fig.\ 10.

\begin{figure}
\centering
\includegraphics[width=0.5\textwidth]{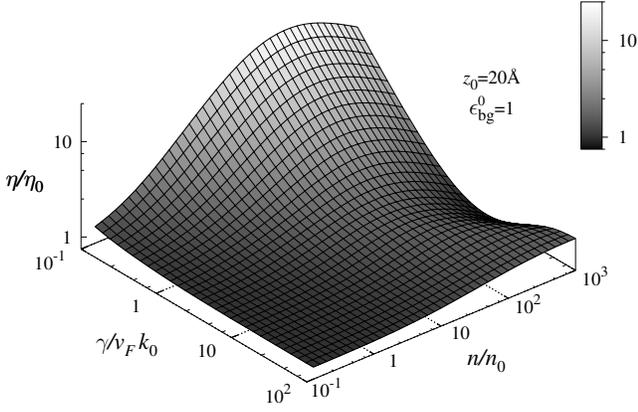}
\caption{The friction coefficient $\eta$ normalized by the friction coefficient for intrinsic graphene $\eta_0$ and shown as a
function of the reduced damping rate $\gamma/(v_Fk_0)$, where $k_0=\sqrt{\pi n_0}$, and the reduced charge-carrier density
$n/n_0$, where $n_0=10^{11}$ cm$^{-2}$, for a proton ($Z=1$) moving at a distance $z_0=20$\AA\ above free graphene
($\ebg^0=1$).}
\end{figure}

\end{section}

\section{Concluding remarks}

We have presented an extensive analysis of the stopping and image forces on an external point charge moving parallel to a single
layer of supported, doped graphene under conditions validating the massless Dirac fermion (MDF) representation of graphene's
$\pi$ electron band excitations. These conditions require that the particle distance, $z_0$, be greater than the lattice
constant of graphene, and that the particle speed, $v$, satisfy $\hbar v/z_0<$ 2 eV. Calculations of the velocity and distance
dependencies of the stopping and image forces were performed within the random phase approximation (RPA) for a broad range of
charge-carrier densities with three major goals: to compare the results with a semi-classical kinetic equation (SCKE) model, to
examine of the effects of a finite graphene-substrate gap, and to explore the effects of a finite damping rate introduced using
Mermin's procedure.

With respect to the first goal, a comparison of the forces from the MDF-RPA and SCKE models in the regime of vanishing damping
has revealed that the latter model may be justified for particle distances satisfying $z_0k_F> 1$ for $v<v_F$ and satisfying
$z_0k_F\gg 1$ for $v>v_F$. When combined through Bohr's adiabatic criterion, these conditions suggest that the SCKE model is
valid only for heavily doped graphene with $\varepsilon_F>\hbar v/z_0$, for which the effects of the inter-band single-particle
excitations on the stopping and image forces are minimized. With respect to the second goal, the effects of a finite gap between
graphene and a supporting substrate in the MDF-RPA model with vanishing damping have been found to be quite strong, particularly
for medium to high particle speeds. These results have confirmed earlier findings from the SCKE model \cite{Radovic_2008} and
raise some concern over the common practice of treating graphene with a zero gap when dealing with the dynamic-polarization
forces on external moving charges. With respect to the third goal, we have made an effort to estimate the order of magnitude of
the damping rate, $\gamma$, by providing a tentative fit of the MDF-RPA dielectric function modified by Mermin's procedure with
recently published experimental data for the HREELS spectra of graphene on a SiC substrate. \cite{Liu_2008} To the best of our
knowledge, this experiment is the only one that has been performed under the parameter constraints of our model. With a suitable
estimate of $\gamma$, we have found that both the stopping and image forces are primarily affected by the damping rate at
moderate particle speeds. We have also shown that the combined effect of a finite gap and a finite damping rate can be important
in modeling the plasmon peak positions of the Mermin loss function.

In all calculations of the stopping and image forces within the MDF-RPA model, we have also found strong effects of the
charge-carrier density, $n$, which are only weakened at sufficiently high particle speeds ($v\gg v_F$). Although both forces
were shown to approach the results for intrinsic graphene in this regime, it is likely that the diminished effect of graphene's
doping level will be masked by excitations of graphene's $\sigma$ electrons as $\hbar v/z_0$ exceeds the $\sigma$ band gap
($\sim$ 7 eV). On the other hand, the parameter ranges considered in this paper are perfectly suitable for studying the friction
of slow charges moving near graphene. The friction coefficient's dependence on the particle distance, the charge-carrier
density, and the damping rate have been studied in detail for a zero gap, where analytic or semi-analytic results have been
obtained. An intricate relationship has been found between these parameters and the friction coefficient, which may be of great
practical interest for applications involving the concept of friction, including surface processing and electrochemistry with
graphene.

There are several possible routes for extending the present work. Although not addressed in this paper, we have examined the
local field correction to the MDF-RPA dielectric function at the level of Hubbard approximation \cite{Adam_2009} and found no
significant effects on the stopping and image forces. However, a recent treatment of the local field effects (LFE) using the
$GW$ method has led to significant improvements in the loss function of free, undoped graphene in the MDF-RPA model with
vanishing damping. \cite{Trevisanutto_2008} Therefore, it would be desirable to treat the LFE using the $GW$ method and explore
its effects on the stopping and image forces. Furthermore, it would be desirable to extend the domain of applicability of the
MDF-RPA model to small particle distances by including the finite size of graphene's $\pi$ orbitals. \cite{Shung_1986} Finally,
the results of our tentative comparison of the Mermin loss function and the experimental HREELS spectra for graphene on a SiC
substrate have opened up the possibility of further modeling of this data in the low frequency regime.

\begin{acknowledgments}
The authors are grateful to N. Bibi\'{c} for insightful discussions and continuing support. K.F.A.\ and Z.L.M.\ acknowledge
support from the Natural Sciences and Engineering Research Council of Canada. The work of D.B., I.R., and Lj.H.\ was supported
by the Ministry of Science and Technological Development, Republic of Serbia.
\end{acknowledgments}

\appendix

\section{The Mermin dielectric function}

In this appendix, we extend the MDF-RPA dielectric function for
graphene given in Refs. \onlinecite{Hwang_2007} and
\onlinecite{Wunsch_2006} to include a finite damping rate
$\gamma>0$. Beginning with Eq.\ (3) of Hwang and Das Sarma,
\cite{Hwang_2007} we rewrite the polarization as
$\Pi(q,\omega+i\gamma)=\Pi^-(q,\omega+i\gamma)+\Pi^+(q,\omega+i\gamma)$,
where it is assumed that $q>0$ and $\omega>0$. For a finite damping
rate, the term $\Pi^-(q,\omega+i\gamma)$ is given by
\begin{equation}
\label{eq:pi_minus}
  \Pi^-(q,\omega+i\gamma)=\frac{g_s g_v q i}{16 \hbar v_F \sqrt{\zeta}},
\end{equation}
where
\begin{equation}
\label{eq:zeta}
  \zeta \equiv \left(\frac{\omega+i\gamma}{v_F q}\right)^2-1.
\end{equation}
Note that the square root of a complex number $z$ is defined as $\sqrt{z}=e^{\frac{1}{2}\log z}$, where $\log z =\log |z|+i \arg
z$ and the argument, $\arg z$, is defined up to an integer multiple of $2\pi$. As with other multi-valued complex functions, the
square root must be evaluated with respect to a single-valued branch, for which the value of $\arg z$ is selected from an
interval of length $2\pi$. For instance, the principal branch, denoted $\text{Arg } z$, selects the value of $\arg z$ from the
interval $(-\pi,\pi]$. In Eq.\ (\ref{eq:pi_minus}), the square root must be evaluated with respect to the branch of $\arg z$
that selects values from $(\theta-2\pi,\theta]$, where $\theta\equiv\text{Arg }\zeta$ is the principal argument of $\zeta$.

The term $\Pi^+(q,\omega+i\gamma)$ is given by
\begin{eqnarray}
\label{eq:pi_plus}
  \Pi^+(q,\omega+i\gamma)=\frac{g_s g_v k_F}{2\pi \hbar v_F}+
  \frac{g_s g_v q}{16 \pi \hbar v_F \sqrt{\zeta}} \times \nonumber \\
\left[
    F\!\left( \frac{\omega+i\gamma}{v_F q} + \frac{2k_F}{ q} \right)
    -
    F\!\left( \frac{\omega+i\gamma}{v_F q} - \frac{2k_F}{ q} \right)
  \right],
\end {eqnarray}
where
\begin{equation}
\label{eq:f}
  F(u)\equiv\frac{u\sqrt{\zeta(u^2-1)}}{\sqrt{\zeta}}-\log\left(u\sqrt{\zeta}+\sqrt{\zeta(u^2-1)}\right)
\end{equation}
and $\zeta$ is defined in Eq.\ (\ref{eq:zeta}). The square roots and logarithms in Eqs.\ (\ref{eq:pi_plus}) and (\ref{eq:f})
must also be evaluated with respect to the branch of $\arg z$ that selects values from $(\theta-2\pi,\theta]$, where
$\theta=\text{Arg }\zeta$. Note that $\sqrt{\zeta(u^2-1)}\ne\sqrt{\zeta}\sqrt{u^2-1}$ when dealing with specific branches of the
square root, so Eq.\ (\ref{eq:f}) cannot be simplified in the obvious manner.

After introducing a finite damping rate $\gamma$, it is necessary to modify the polarization $\Pi(q,\omega+i\gamma)$ using
Mermin's procedure to conserve the local number of charge carriers in graphene.\cite{Mermin_1970, Qaiumzadeh_2008} The Mermin
polarization function is then given by
\begin{equation}
\label{Pi_M} \Pi_M(q,\omega,\gamma)=\frac{\Pi(q,\omega+i\gamma)}{1-\displaystyle{\frac{i\gamma}{\omega+i\gamma}\left[1-
\frac{\Pi(q,\omega+i\gamma)}{\Pi_s(q)}\right]}},
\end{equation}
where $\Pi_s(q)\equiv\Pi(q,0)$ is the static limit of the polarization $\Pi(q,\omega+i\gamma)$ with $\omega\rightarrow0$ and
$\gamma\rightarrow0$, given by \cite{Wunsch_2006,Hwang_2007}
\begin{eqnarray}
\label{eq:Pi_s} \Pi_s(q) = \frac{g_sg_vk_F}{2\pi \hbar v_F} \times
\nonumber \\
\left\{
 \begin{array}{ll}
 1 & \text{if }q \le 2k_F \\
  1-\frac{1}{2}\sqrt{1-\frac{4k_F^2}{q^2}}-\frac{q}{4k_F}\arcsin{\left(\frac{2k_F}{q}\right) +\frac{\pi q}{8k_F}} &
  \text{otherwise} \\
\end{array}
\right\}.
\end{eqnarray}

\noindent Note that the piecewise definition of Eq.\ (\ref{eq:Pi_s})
ensures that the square root and arcsine are real-valued. Using Eq.\
(\ref{Pi_M}) in Eq.\ (\ref{eps}), the Mermin dielectric function is
then given by
\begin{equation}
\label{eq:e_M} \epsilon_M(q,\omega,\gamma)=\ebg(q)+\frac{2\pi e^2}{q}\Pi_M(q,\omega,\gamma).
\end{equation}

In the limit $\gamma \rightarrow 0$, it can be shown that Eq.\ (\ref{eq:e_M}) is equivalent to the dielectric functions
presented in Refs.\cite{Wunsch_2006,Hwang_2007} \cite{Note} The difficulty in obtaining a compact form for the dielectric
function in the limit as $\gamma\rightarrow0$ lies in reproducing the behaviour of the branch cut that naturally arises for
$\gamma>0$ --- it is necessary to employ rather complicated piecewise-defined functions, as in
Refs.\cite{Hwang_2007,Wunsch_2006} Alternatively, the compact Eq.\ (\ref{eq:e_M}) may be used with a small, positive $\gamma$ to
approximate this limit, or Eq.\ (\ref{eq:e_M}) may be used with a realistic damping rate, $\gamma$, as originally intended. To
this end, we describe how to implement the branch cut technique below.

Most computer codes support basic arithmetic for complex numbers
including exponentiation with base $e$, but only include built-in
functions for computing complex logarithms and square roots with
respect to the principal branch. To compute logarithms and square
roots with respect to the branch required in Eqs.\
(\ref{eq:pi_minus}), (\ref{eq:pi_plus}), and (\ref{eq:f}), it is
necessary to implement a function \texttt{argbranch(z)} that returns
the argument of a complex number \texttt{z} in the range
$(\theta-2\pi,\theta]$, where $\theta=\text{Arg }\zeta$. Using the
built-in function $\texttt{atan2(y,x)}$, which returns the polar
angle of the Cartesian point (x,y) in the range $(-\pi,\pi]$, one
may define
\begin{equation}
\label{eq:Arg}
  \texttt{Arg(z)} : \texttt{z} \mapsto \texttt{atan2(Im(z),Re(z)),}
\end{equation}
\begin{eqnarray}
\label{eq:argbr} \texttt{argbranch(z)} : \nonumber \\
\texttt{z}
\mapsto \left\{
  \begin{array}{ll}
    \texttt{Arg(z)}-\texttt{2}\pi & \text{if } \texttt{Arg(z)} > \texttt{Arg($\zeta$)} \\
    \texttt{Arg(z)} & \text{otherwise}
  \end{array}
\right..
\end{eqnarray}
Since $q>0$ and $\omega>0$, Eq.\ (\ref{eq:argbr}) correctly handles
all values of \texttt{z}. The functions for the corresponding
branches of the logarithm and the square root are then defined as
\begin{equation}
\label{eq:logbr}
  \texttt{logbranch(z)} : \texttt{z} \mapsto \texttt{log(|z|)}
  + i \texttt{argbranch(z)},
\end{equation}
\begin{equation}
\label{eq:sqrtbr}
  \texttt{sqrtbranch(z)} : \texttt{z} \mapsto
  \texttt{exp(logbranch(z)/2)},
\end{equation}
where \texttt{|z|} is the modulus of \texttt{z}, \texttt{log()} and \texttt{sqrt()} are the real-valued logarithm and square
root, respectively, and \texttt{exp()} is the complex exponentiation with base $e$. It is then a simple matter of evaluating
Eqs.\ (\ref{eq:pi_minus}), (\ref{eq:pi_plus}), and (\ref{eq:f}) with respect to these branches of the logarithm and square root.
Note that \texttt{argbranch($\zeta$)} must return \texttt{Arg($\zeta$)} to yield the correct result.

\end{document}